\documentclass[dvipdfmx,10pt,aps,twocolumn,prc,superscriptaddress,showpacs,nofootinbib,noshowkeys,floatfix,preprintnumbers,reprint]{revtex4}
\usepackage[dvipdfmx]{graphicx}
\usepackage{color}
\usepackage{amsmath, amssymb}
\usepackage{multirow}
\usepackage{longtable}
\usepackage[normalem]{ulem}
\usepackage{epstopdf}
\usepackage[normalem]{ulem}  

\renewcommand\sout{\bgroup \color{red} \ULdepth=-.5ex \ULset}
\begin{document}
\title{Probing Multi-Strange Dibaryon with Proton-Omega Correlation \\ in High-energy Heavy Ion Collisions}
\date{\today}
\author{Kenji Morita}
\email{kmorita@yukawa.kyoto-u.ac.jp}
\author{Akira Ohnishi}
\email{ohnishi@yukawa.kyoto-u.ac.jp}
\affiliation{Yukawa Institute for Theoretical Physics, Kyoto University,
Kyoto 606-8502, Japan}
\author{Faisal Etminan}
\email{fetminan@birjand.ac.ir}
\affiliation{Department of Physics, Faculty of Sciences, University of Birjand, 
Birjand 97175-615, Iran}
\author{Tetsuo Hatsuda}
\email{thatsuda@riken.jp}
\affiliation{Theoretical Research Division, Nishina Center, RIKEN,
Saitama 351-0198, Japan\\
iTHES Research Group, RIKEN, Saitama 351-0198, Japan }
\preprint{YITP-16-62,RIKEN-QHP-223}
\begin{abstract}
The two-particle momentum correlation
between the proton ($p$) and the Omega-baryon ($\Omega$) 
in high-energy heavy ion collisions is studied to 
unravel the possible spin-2 $p\Omega$ dibaryon  recently 
suggested by lattice QCD simulations.
The ratio of correlation functions between small and large collision systems,
$C_{\rm SL}(Q)$, is proposed to be a  new measure to extract the strong $p\Omega$ interaction 
without much contamination from the Coulomb attraction.  Relevance of this quantity to the 
experimental observables in heavy-ion collisions is also discussed. 
\end{abstract}
\maketitle

\textit{Introduction.\textemdash}
Baryon-baryon interactions serve as crucial inputs for understanding
possible dibaryons with strangeness ($S$). In particular, the spin-0 $H$
state with $S=-2$ \cite{jaffe77:_perhap} and the spin-2 nucleon-Omega
($N\Omega$) state with $S=-3$ \cite{goldman87:_stran} are among the most
promising candidates for bound or resonant dibaryons, since the Pauli
exclusion principle does not operate among quarks in these channels
\cite{oka88:_flavor,Gal:2015rev}. 
Indeed, it was recently reported from first-principles lattice QCD
simulations with heavy quark masses that there exist sizable attractions
in the spin-0 $H$ channel \cite{Inoue:2010hs,Beane:2010hg} and in the
spin-2 $N\Omega$ channel \cite{Etminan:2014tya},
although the quantitative answers would be obtained only by the on-going physical
point simulations \cite{Doi:2015oha}.

Experimentally, high-energy heavy ion collisions at the BNL Relativistic
Heavy-Ion Collider (RHIC) and the CERN Large Hadron Collider (LHC) provide
unique opportunities for multi-strange dibaryons \cite{Cho:2010db}.
For example, the search for the $H$-dibaryon
 has been conducted  both at RHIC
\cite{adamczyk15:_LL} and LHC \cite{adam16:_searc_lambd_pb_pb_nn}.
More generally, the final state interactions for identical and
non-identical hadrons after freeze-out \cite{lednicky82:_influence,Bauer:1993wq}
have been shown to be sensitive to the low-energy scattering parameters
\cite{adamczyk15:_antiprotonc2,Adam:2015vja,adamczyk15:_LL,Morita:2014kza,Ohnishi:2016elb,Shapoval:2014yha,Kerbikov:2009ym}.

In this paper, motivated by the recent study on the  $N\Omega$ interaction
 in lattice QCD \cite{Etminan:2014tya}, we
study the proton-omega ($p\Omega$) correlation function in the relativistic
heavy ion collisions to probe the nature of the $S=-3$
dibaryon. We propose that the ratio of the $p\Omega$ correlation functions
 with different source sizes is a  good measure to extract the
 strong interaction between  $p$ and $\Omega$  without much contamination from the 
 Coulomb attraction. 

\textit{$N\Omega$ interaction.}\textemdash 
In the S-wave $N\Omega$ system, there exist two possible channels,
$^5{\rm S}_2$ and $^3{\rm S}_1$, where $ ^{2s+1}L_J$ denotes  the state
with spin-$s$, $L$-wave, and total angular momentum $J$. 
As long as the strong interaction is concerned,
 the lowest threshold in octet-decuplet and decuplet-decuplet channels 
 is $N\Omega$ at 2610MeV. In the octet-octet channel, 
 $\Lambda \Xi$ (with the threshold at 2430MeV) and $\Sigma \Xi$ (at 2507MeV)
 lie below $N\Omega$.
  For $^5{\rm S}_2$, the coupling of $N\Omega$ to these octet-octet channels
occurs only through the D-wave, and thus it is dynamically suppressed. On
the other hand, for $^3{\rm S}_1$,  the sizable coupling to
octet-octet channels through the rearrangement of quarks
is expected  in the S-wave.

\begin{figure}[!tb]
 \centering
 \includegraphics[width=3.2in]{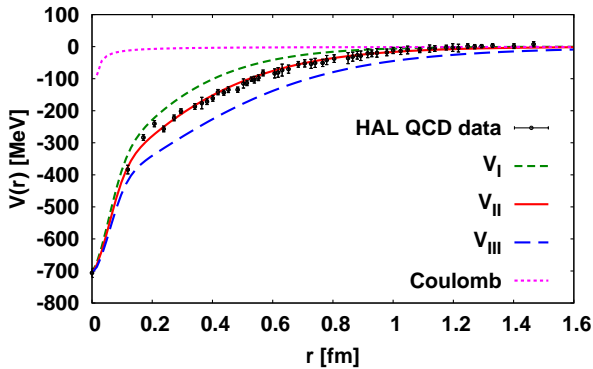}
 \caption{Three typical examples of the $N\Omega$ potential. 
 The black circles with error bars  stand for the lattice QCD data 
  with heavy quark masses \cite{Etminan:2014tya}.
 The red line ($V_{\rm II}$) corresponds to a fit to the lattice 
 data with a Gaussian + (Yukawa)$^2$ form. 
 The green short-dashed line ($V_{\rm II}$) and the blue long-dashed line ($V_{\rm III}$)
  denote the potentials with weaker and stronger attractions, respectively. 
 The Coulomb potential for the $p\Omega$ system is also shown by the purple dashed line. }
 \label{fig:potential_mphys}
\end{figure}

Let us first consider the $N\Omega$ interaction in the $^5{\rm S}_2$ channel where
 recent lattice QCD simulations with heavy quarks
($m_{\pi}$=875 MeV and $m_K$=916 MeV) show attraction
 for the entire range of their relative distance $r$ \cite{Etminan:2014tya}.
In Fig.~\ref{fig:potential_mphys}, the lattice data are shown by
  black circles with statistical
error bars. The data can be fitted well by
 an attractive Gaussian core + an attractive (Yukawa)$^2$ tail with a form factor;
$ V(r) = b_1 e^{-b_2 r^2} + b_3 (1-e^{-b_4 r^2}) (e^{-b_5r}/r)^2$,
where $b_{1,3}<0$ and $b_{2,4,5}>0$. The best fit is shown by the red solid
line in the figure denoted by $V_{\rm II}$. 
Assuming that the qualitative form of this attractive potential does not
change even for physical quark masses, we generate a series of potentials 
 by varying the range-parameter at long distance, $b_5$.
Two typical examples are $V_{\rm I}$ with weaker attraction 
and $V_{\rm III}$ with stronger attraction in Fig.~\ref{fig:potential_mphys}. 
By solving the Schr\"{o}dinger equation using these potentials with the 
physical baryon masses, one finds no bound state for $V_{\rm I}$,
 a shallow bound state for $V_{\rm II}$, and a deep bound state for $V_{\rm III}$.
  The binding energies, the scattering lengths
\footnote{We use the ``nuclear physics convention'' in which the scattering
phase shift $\delta$ at small momentum is given as $\delta = -ka_0$.}
and the effective ranges in the $^5{\rm S}_2$ $p\Omega$ channel with and
without the Coulomb potential are summarized in Table \ref{tbl:pot}. 

As for the $N\Omega$ interaction in the $^3{\rm S}_1$ channel, there would be
a strong coupling to the octet-octet channels when the spatial distance between
$N$ and $\Omega$ becomes small.  We model this by complete
absorption of the $N\Omega$ wave function at short distance $r < r_0$.  
In other words, $V(r;\ ^3S_1)= -i V_0 \theta (r_0-r) $ with $V_0 \rightarrow + \infty$ for the 
strong interaction part.
We  choose $r_0=2$ fm, because the Coulomb potential dominates over
$V_{\rm I,II,III}$ for $r> 2$ fm.

\textit{$N\Omega$ Correlation function.\textemdash}
 The pairwise hadronic interaction gives enhancement or reduction of the 
 observed number of pairs with small relative momenta in heavy ion collisions.
 In particular, the correlation function of non-identical pairs
  is  directly related to the pairwise interaction due to the 
 absence of the quantum statistical effect \cite{Bauer:1993wq}.
The $p\Omega$ correlation function is given in terms of the two-particle distribution
$N_{p\Omega}(\boldsymbol{k}_p,\boldsymbol{k}_\Omega)$ normalized by the
product of the single particle distributions,
$N_\Omega(\boldsymbol{k}_\Omega) N_p(\boldsymbol{k}_p)$, with
$\int_{x_i}\equiv \int d^4 x_i$,
 \begin{eqnarray}
 C(\boldsymbol{Q},\boldsymbol{K}) &=&
  \frac{N_{p\Omega}(\boldsymbol{k}_p,\boldsymbol{k}_\Omega)}{N_p(\boldsymbol{k}_p)
  N_\Omega(\boldsymbol{k}_\Omega)}  \label{eq:full-CQ}
   \\ 
  &\simeq& \frac{ \int_{x_p} \int_{x_\Omega} S_p(x_p, \boldsymbol{k}_p)
  S_\Omega(x_\Omega, \boldsymbol{k}_\Omega) 
  \left| \Psi_{p\Omega}(\boldsymbol{r}^\prime)\right|^2}
  {\int_{x_p}S_p(x_p,\boldsymbol{k}_p) \int_{x_\Omega} S_\Omega(x_\Omega, \boldsymbol{k}_\Omega)},
 \nonumber
\end{eqnarray}
where relative and total momenta are defined as
$ \boldsymbol{Q} = (m_p \boldsymbol{k}_\Omega - m_\Omega \boldsymbol{k}_p)/M$
 and  $ \boldsymbol{K} = \boldsymbol{k}_p + \boldsymbol{k}_\Omega$, respectively,  
with $M\equiv m_p + m_\Omega$.
The source functions 
$S_i(x_i, \boldsymbol{k}_i)\equiv E_i \frac{dN_i}{d^3\boldsymbol{k}_i d^4 x_i} $, with $i=p, \Omega$
 and $E_i= \sqrt{\boldsymbol{k}_i^2+m_i^2}$,  denote the phase space distribution
of $p$ and $\Omega$ at freeze-out.  The final state interaction
after the freeze-out is described by the two-particle wave
function $\Psi_{p\Omega}$, in which the shift of the relative coordinate
$\boldsymbol{r}=\boldsymbol{x}_\Omega-\boldsymbol{x}_p$ to
$\boldsymbol{r}^\prime = \boldsymbol{r}-\boldsymbol{K}(t_p-t_\Omega)/M$
accounts for the possible difference in the emission time between $p$
and $\Omega$.
In the following, we assume that the pair purity is unity; 
i.e.  the weak decay contribution to $p$ can be removed experimentally 
and  that to $\Omega$ is negligible
~\cite{adamczyk15:_LL,Agakishiev:2011ar}.

\begin{table}[!tb]
 \caption{The binding energy ($E_{\rm B}$), 
 the scattering length ($a_0$) and the effective range ($r_{\rm eff}$) with and without 
 the Coulomb attraction in the $p\Omega$ system. Physical masses of the proton and 
 $\Omega$ are used.}
\begin{tabular}[t]{lc|ccc}\hline
 \multicolumn{2}{c|}{Spin-2 $p\Omega$ potentials}  & $V_{\text{I}}$ &$V_{\text{II}}$&$V_{\text{III}}$\\\hline
 & $E_{\rm B}$~[MeV] &$-$& 0.05 &24.8 \\
without Coulomb & $a_0$~[fm]& $-1.0$ & 23.1&1.60 \\
                       & $r_{\text{eff}}$~[fm]&1.15&0.95&0.65 \\ \hline
   & $E_{\rm B}$~[MeV]&$-$&6.3 & 26.9  \\
with Coulomb  &$a_0$~[fm]&$-1.12$ &5.79&1.29 \\
                       &$r_{\text{eff}}$~[fm]&1.16&0.96 &0.65 \\ \hline
\end{tabular}
 \label{tbl:pot}
\end{table}

Taking into account the spin degeneracy, we have  
$|\Psi_{p\Omega} |^2 = \frac{5}{8} |\Psi_5(\boldsymbol{r})|^2 + \frac{3}{8} |\Psi_3(\boldsymbol{r})|^2$,
where  $\Psi_5$ ($\Psi_3$) denotes the wave functions in spin-2 (spin-1) channel.
The strong interaction is short ranged and modifies only the S-wave
component of the wave function, so that we may write
\begin{eqnarray}
 \Psi_{5(3)}(\boldsymbol{r}) =   [\psi^C(\boldsymbol{r}) - \psi_{0}^C(r)] + \chi_{\rm sc(abs)}(r).
\label{eq:P53}
 \end{eqnarray}
Here $\psi^C(\boldsymbol{r})$ is the Coulomb wave function
 characterized by the reduced mass $\mu=601$ MeV and the Bohr radius 
$a_{\rm B} = (\mu \alpha)^{-1} \simeq 45$ fm of the $p\Omega$ system.
Its S-wave component is denoted by $\psi_{0}^C(r)$.
The scattering wave function in the $^5{\rm S}_2$ state,  $\chi_{\rm sc}(r)$, is obtained by solving
the Schr\"{o}dinger equation with both the strong interaction ($V_{\rm I, II, III}$) and the Coulomb interaction. 

Note that $\chi_{\rm sc}(r)$ reduces to $\psi_{0}^C(r)$ in the absence of strong interaction.
On the other hand, the wave function $\chi_{\rm abs}(r)$ in the $^3{\rm S}_1$ channel
 vanishes for $r\le  r_0$ due to our assumption of the complete absorption into octet-octet states
\footnote{
For complete absorption ($V_0 \rightarrow +\infty$), the scattering length without 
the Coulomb interaction in the $~^3S_1$  channel becomes (Re$\ a_0$, Im$\ a_0$)=($r_0, 0)$,
 which is equivalent to the hard sphere with a radius
$r_0$. For finite  $V_0$, one has Re$\ a_0 < r_0$ and Im$\ a_0 < 0$
(see \cite{Shapoval:2014yha} for the analysis of  
 baryon-antibaryon correlation function).  We have checked that finite $V_0$
  leads to a reduction of  $C(Q)$ in the  $~^3S_1$ channel particularly for $Q< 50$ MeV,
   but the effect on the total $C(Q)$ is small, so that our conclusions are unchanged.} 
, while  it is identical to the Coulomb wave function for $r > r_0$:
\begin{equation}
 \chi_{\text{abs}}(r)  = \theta(r-r_0)  \frac{1}{2i\bar{r}}
  \left( H_0^+(\bar{r}) - F(\bar{r}_0) H_0^-(\bar{r}) \right).
\end{equation}
Here $Q=|\boldsymbol{Q}|$, $\bar{r}=Qr$, $\bar{r}_0=Qr_0$, and  
 $H_{L=0}^{+}$ ($H_{L=0}^{-}$) is the 
outgoing (incoming) Coulomb function which reduces to $e^{+i\bar{r}}$ ($e^{-i\bar{r}}$) 
without the Coulomb force \cite{Handbook_math}. 
Note that $F(\bar{r}_0)=H_0^{+}(\bar{r}_0)/ H_0^{-}(\bar{r}_0)$,
so that $\chi_{\text{abs}}(r)$ is continuous across $r=r_0$. 
In the absence of the absorption, we have
$\left. \chi_{\text{abs}}(r)\right|_{r_0 \rightarrow 0} =
\psi^C_{0}(r)$, since $F(\bar{r}_0=0)=1$.

\textit{Case with static source.\textemdash}
We now consider the following static source function with spherical symmetry
to extract the essential part of physics;
\begin{equation}
 S_i(x_i,\boldsymbol{k}_i) 
 = {\cal N}_i  E_i \ e^{-\frac{\boldsymbol{x}_i^2}{2R_i^2}} \ \delta(t-t_i), \ \ (i=p, \Omega).
 \label{eq:staticsource}
\end{equation}
Here $R_i$ is a source size parameter, while 
${\cal N}_i$ is a normalization factor which cancels out between the 
numerator and denominator together with $E_i$ in Eq.(\ref{eq:full-CQ}).
Assuming the equal-time emission $t_p=t_\Omega$ for the moment, 
one obtains a concise formula,
 \begin{eqnarray}
C(Q) & =&  \int [dr] \int \frac{d\Omega}{4\pi}  |\psi^C(\boldsymbol{r})|^2 
   \nonumber \\
 &+& \frac{5}{8}  \int [dr]
   \{ |\chi_{\rm sc}(r)|^2 - |\psi_{0}^C(r)|^2 \} \nonumber\label{} \\   
 &+& \frac{3}{8} \int [dr]
   \{ |\chi_{\text{abs}}(r)|^2 - |\psi_{0}^C(r)|^2 \} ,
 \label{eq:c2static}
\end{eqnarray}
where $[dr]=\frac{1}{2\sqrt{\pi}R^3} \! dr \, r^2  e^{-\frac{r^2}{4R^2}}$
 with $R = \sqrt{(R_p^2 + R_\Omega^2)/2}$ being the effective size
parameter. $\int d\Omega$ is the integration over the solid angle between
$\boldsymbol{Q}$ and $\boldsymbol{r}$. 
Without the Coulomb interaction, the integration of the first line in Eq.(\ref{eq:c2static})
simply gives unity.
 The second (third) line is nothing but a spatial average of the difference between the S-wave 
 probability density with and without the strong interaction, where the source function acts as a weight factor.
A similar formula for the
$\Lambda\Lambda$ correlation
has been previously derived  with the quantum statistical effect \cite{Morita:2014kza}.

\begin{figure}[!tb]
 \centering
 \includegraphics[width=3.2in]{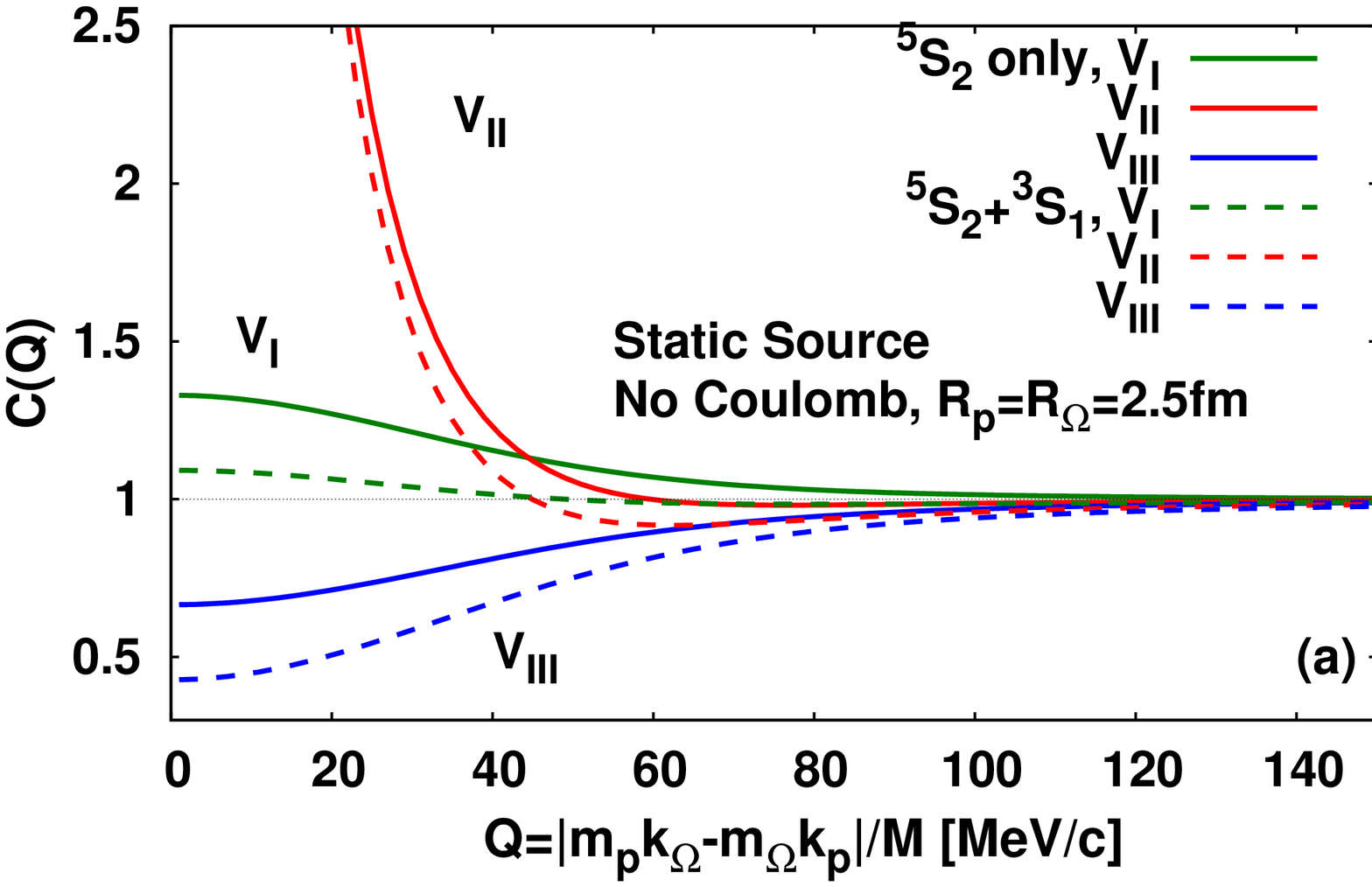}
 \includegraphics[width=3.2in]{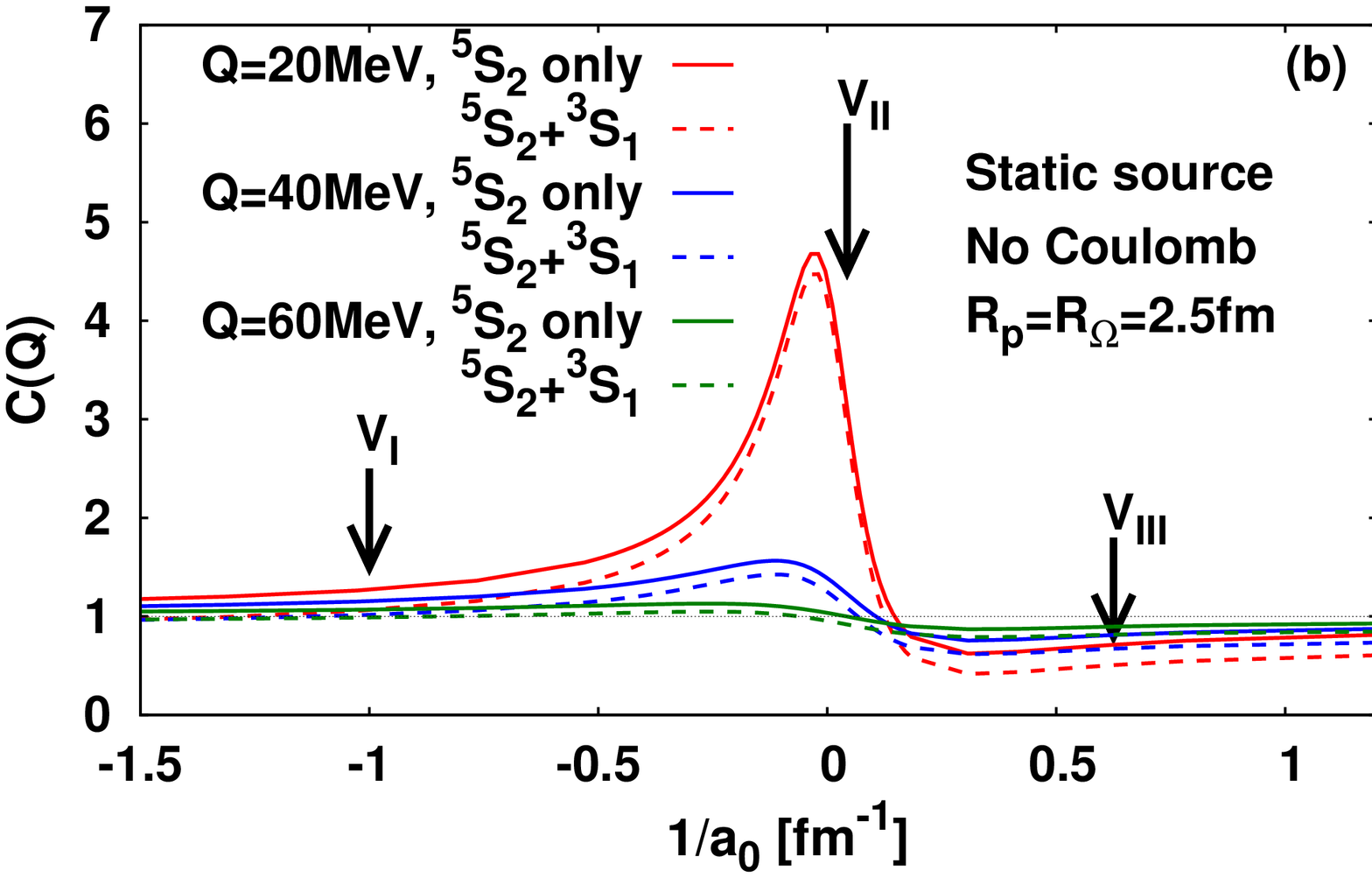}
 \caption{$p\Omega$ correlation function for the static source with 
$R_p=R_\Omega=2.5$~fm.
 The Coulomb interaction is switched off.  (a) Solid (dashed) lines
 denote the correlations  with only the $^5{\rm S}_2$ scattering 
  (with both the $^5{\rm S}_2$ scattering and the $^3{\rm S}_1$ absorption).
 (b) $C(Q)$ for $Q=20 $, 40 and 60 MeV as a function of $a_0^{-1}$ obtained by changing the 
 attraction of the $N\Omega$ potential through the parameter $b_5$.}
 \label{fig:c2}
\end{figure}

Let us now discuss,  
step by step on the basis of Eq.(\ref{eq:c2static}),
 the effects of the elastic scattering in the $^5{\rm S}_2$ channel, 
the strong absorption in the $^3{\rm S}_1$ channel and the long range Coulomb interaction.  
First,  we neglect the Coulomb interaction,
so that the first line of Eq.~\eqref{eq:c2static} is unity and $\psi_{0}^C(r)$ becomes
the free spherical wave $ j_0(\bar{r})$.
For a weak attraction without bound state ($V_{\rm I}$), the probability density
$|\chi_{\rm sc}(r)|^2$  in the $^5{\rm S}_2$ channel is slightly enhanced from the free one  at short
distances and at small $Q$.  This  leads to $C(Q)$ represented by the
green solid curve in Fig.~\ref{fig:c2}(a) illustrated for a characteristic
source size $R_p=R_\Omega =2.5$ fm measured in the $pp$ correlation for mid-central events
\cite{adamczyk15:_antiprotonc2,Adam:2015vja}.
 As the attraction increases
towards and across the unitary limit where the scattering length
diverges, the enhancement of $C(Q)$ becomes prominent as represented by
the red solid curve corresponding to $V_{\rm II}$ in
Fig.~\ref{fig:c2}(a). As the attraction becomes even stronger,
$\chi_{\rm sc}(r)$ for small $Q$ starts to oscillate and to be 
suppressed inside the  source radius $R$ due to large local momentum 
$q(r)=\sqrt{2\mu(E-V)}$. This
effect tames the enhancement of $C(Q)$ and eventually leads to a 
suppression of $C(Q)$ as represented by  the blue solid curve
corresponding to $V_{\rm III}$ in Fig.~\ref{fig:c2}(a).

In the $^3{\rm S}_1$ channel, the probability
density $|\chi_{\rm abs}(r)|^2$ is zero at short distances. This implies that
the absorption effect tends to suppress the particle correlation as 
indicated by the third line of Eq.~\eqref{eq:c2static}.  
The dashed lines in
Fig.~\ref{fig:c2}(a) show $C(Q)$ with both the $^5{\rm S}_2$ scattering and
the  $^3{\rm S}_1$ absorption: The absorption effect is not negligibly small,
but is not significantly large enough to change the qualitative behavior of
$C(Q)$ obtained by the $^5{\rm S}_2$ scattering alone.

Shown in Fig.~\ref{fig:c2}(b) is $C(Q)$ without the Coulomb interaction
for three typical momenta ($Q=20$,40, and 60 MeV) as a function of  $a_0^{-1}$. 
 By shifting the parameter $b_5$ in the $N\Omega$ potential,
 one can change the scattering length $a_0$ from negative to positive values without
 substantial change of the effective range $r_{\text{eff}}$:
 The arrows in the figure indicate  $a_0^{-1}$ corresponding to $V_{\rm I,II,III}$.
For weak (strong) attraction where $a_0^{-1} < 0$ $(a_0^{-1}> 0)$, 
$C(Q)$ is enhanced  (suppressed) from unity, while it is substantially enhanced around the 
unitary limit $a_0^{-1} =0$.  This implies that $C(Q)$ for a certain range of $Q$
would provide a useful measure to identify the strength of the $N\Omega$ attraction.

\begin{figure}[!tb]
 \centering
 \includegraphics[width=3.2in]{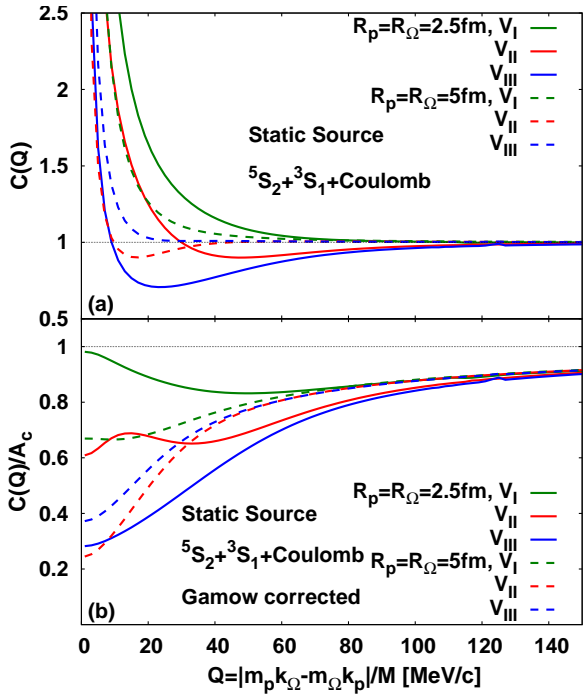}
 \caption{(a) Correlation function  with both strong and the Coulomb attractions
 for two different values of the static source sizes, $R_{p,\Omega}=2.5$~fm (solid lines) and 5~fm 
 (dashed lines).   (b) Same correlation function as (a), but divided by the Gamow factor.}
 \label{fig:c3}
\end{figure}

Once we include the Coulomb interaction between the positively charged $p$ and 
the negatively charged $\Omega$, 
a strong enhancement of $C(Q)$  at small $Q$ is introduced by the long-range attraction.  
The results with the Coulomb attraction are shown by the solid lines in
Fig.~\ref{fig:c3}(a) for $R_{p,\Omega}=2.5$~fm. 
One finds that (i) the difference among three curves with $V_{\rm I, II, III}$ in
Fig.~\ref{fig:c2}(a) is less visible in Fig.~\ref{fig:c3}(a)  at small $Q$ due to the Coulomb enhancement,
and (ii) the ordering of three curves become different especially due to the 
large  reduction of the scattering length for $V_{\rm II}$ by  the Coulomb effect (Table
\ref{tbl:pot}).
The dashed lines in  Fig.~\ref{fig:c3}(a) represent $C(Q)$ for larger
source size,  $R_{p,\Omega}=5$~fm. In this case, the correlation function is
more sensitive to the long-range part of the interaction as  found for
the proton-proton correlation \cite{lednicky82:_influence,Bauer:1993wq}. 
As a result, the ordering of the correlation function is further changed
such that $C(Q)$ for $V_{\text{II}}$ becomes
the lowest.\footnote{For large $R$, the integrals in Eq.~\eqref{eq:c2static} are
dominated by contributions from the outside of the potential range,
where the wave functions are solely determined by the scattering phase
shift. Then the effective range approximation for the S-wave scattering
length leads to the Lednick\'{y}-Lyuboshitz formula
\cite{lednicky82:_influence}, in which $C(Q)$ is not sensitive to the
potential shape \cite{Gmitro:1986ay} and is expressed in terms of 
low-energy scattering parameters.}

\begin{figure}[!tbh]
 \includegraphics[width=3.2in]{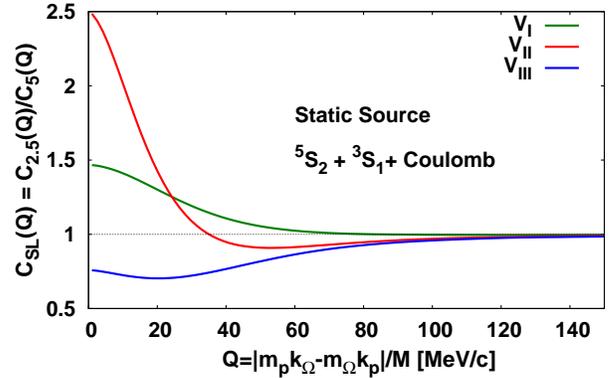}
 \caption{$C_{\rm SL}(Q)$ for the static source
 between the different source sizes, $R_{p,\Omega}=2.5$ and 5 fm.}
 \label{fig:csl}
\end{figure}

One may try to remove the  Coulomb enhancement  in Fig.\ref{fig:c3}(a)
  by dividing $C(Q)$ by the $R_{p,\Omega}$-independent Gamow factor, $A_c(\eta)=2\pi\eta/(e^{2\pi\eta}-1)$,
 with $\eta=-(Q a_{\rm B})^{-1}$ being the Sommerfeld parameter.
Comparison of  $C(Q)/A_c$ in  Figure \ref{fig:c3}(b) and $C(Q)$ in Figure \ref{fig:c2}(a)
 indicates that the simple Gamow correction is  not good enough to extract the  characteristic feature
of $C(Q)$ from the strong interaction:  In principle, full Coulomb
correction with source-size dependence is needed to  isolate the effect of strong interaction.
As an alternative and model-independent way to handle the Coulomb 
effect,  we propose to introduce an ``SL (small-to-large) ratio'' of the correlation functions
for systems with different source sizes,
\begin{eqnarray}
C_{\rm SL}(Q)\equiv \frac{C_{R_{p,\Omega}=2.5{\rm fm}}(Q)}{C_{R_{p,\Omega}=5{\rm fm}}(Q)},
\end{eqnarray}
as  shown in Figure \ref{fig:csl}.
An advantage of this ratio is that the effect of the Coulomb interaction 
for small $Q$ is largely canceled, so that it has a good 
sensitivity to the strong interaction without much contamination from the 
Coulomb interaction.

\textit{Effects of expansion and freeze-out time.\textemdash}
The results so far have been obtained with a simplified static
source function \eqref{eq:staticsource}. In reality, 
the collective expansion takes place in high energy heavy ion collisions.
Also, the freeze-out of multi-strange hadrons may occur
prior to other hadrons due to small cross sections
\cite{zhu15:_hybrid,takeuchi15:_effec}. 
To see the influences of these dynamical properties, we consider the
following source model with a 1-dim Bjorken expansion
 \cite{Chapman:1994ax},
\begin{equation}
 S(x_i,\boldsymbol{k}_i) = {\cal N}_i' E^{\rm tr}_i
  \frac{1}{e^{E^{\rm tr}_i /T_i}+1}e^{-\frac{x^2+y^2}{2(R_i^{\rm tr})^2}}\ 
  \delta(\tau-\tau_i),\label{eq:expandingsource}
\end{equation}
where $E^{\rm tr}_i=\sqrt{(\boldsymbol{k}_i^{\rm{tr}})^2+m_i^2}  \cosh(y_i-\eta_s)$ with
 the momentum rapidity $y_i$ and the space-time rapidity 
 $\eta_s = \ln\sqrt{(t+z)/(t-z)}$.  
The temperature and the proper-time at the 
 thermal freeze-out are denoted by $T_i$ and  $\tau_i$, respectively.
 The transverse source size is denoted 
 by the parameter $R_i^{\rm tr}$.
We have not taken into account the transverse collective expansion explicitly in the present paper,
since its effect  on $C(Q)$ has been shown to be  
effectively absorbed into a slight modification of $R_i^{\rm  tr}$  as shown 
for  the $\Lambda\Lambda$ correlation with the same model \cite{Morita:2014kza}.

We consider a small system with $R_p^{\rm tr} = R_\Omega^{\rm tr}=2.5$~fm
and a large system with $R_p^{\rm tr}=R_\Omega^{\rm tr}=5$~fm.
Following the results of the dynamical analyses of the peripheral and
 central Pb+Pb collisions at $\sqrt{s_{NN}}=2.76$ TeV with hydrodynamics
 + hadronic transport \cite{zhu15:_hybrid}, we take  $\tau_p \ (\tau_\Omega)  =3 \ (2)$~fm for the
former, and $\tau_p \ (\tau_\Omega)=20\ (10)$~fm for the latter as
characteristic values.
We take $T_{p,\Omega}=$164~MeV  for peripheral collisions \cite{ABELEV:2013zaa}, while
 $T_p (T_{\Omega})$=120\ (164) MeV for central collisions \cite{Shen:2011eg}.
Under the expanding source, Eq.(\ref{eq:full-CQ}) has explicit $\boldsymbol{K}$ dependence:
For illustrative purpose, we take the total longitudinal momentum 
 to be zero $K_z=0$ and the total transverse momentum to be
$|\boldsymbol{K}^{\rm  tr}|$=2.0\ (2.5) GeV for peripheral (central) collisions 
 which correspond to the twice the mean
$|\boldsymbol{k}_p^{\rm tr}|$ values of the proton \cite{Abelev:2013vea}.

\begin{figure}[!tb]
 \centering
  \includegraphics[width=3.2in]{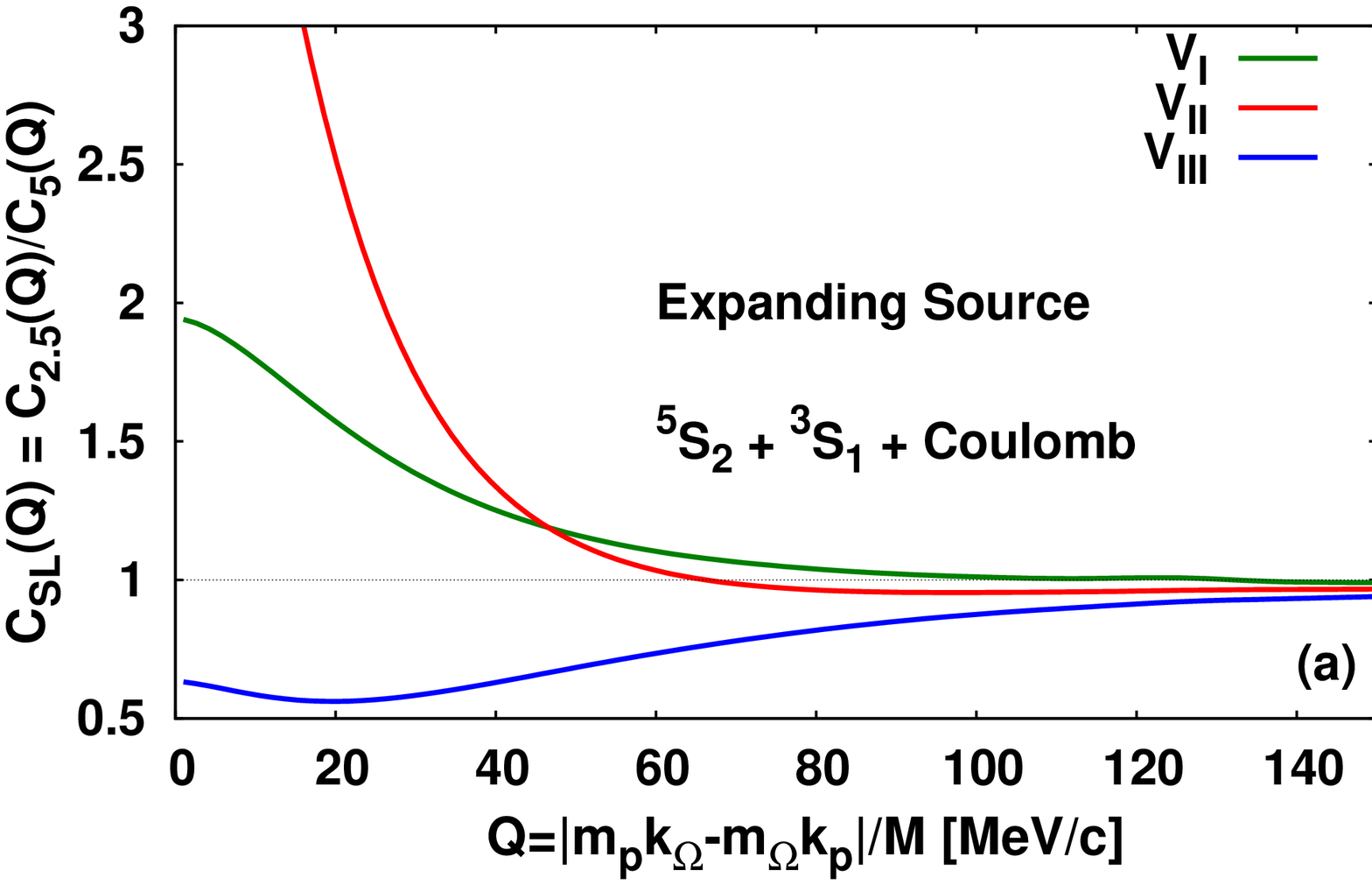}
  \includegraphics[width=3.2in]{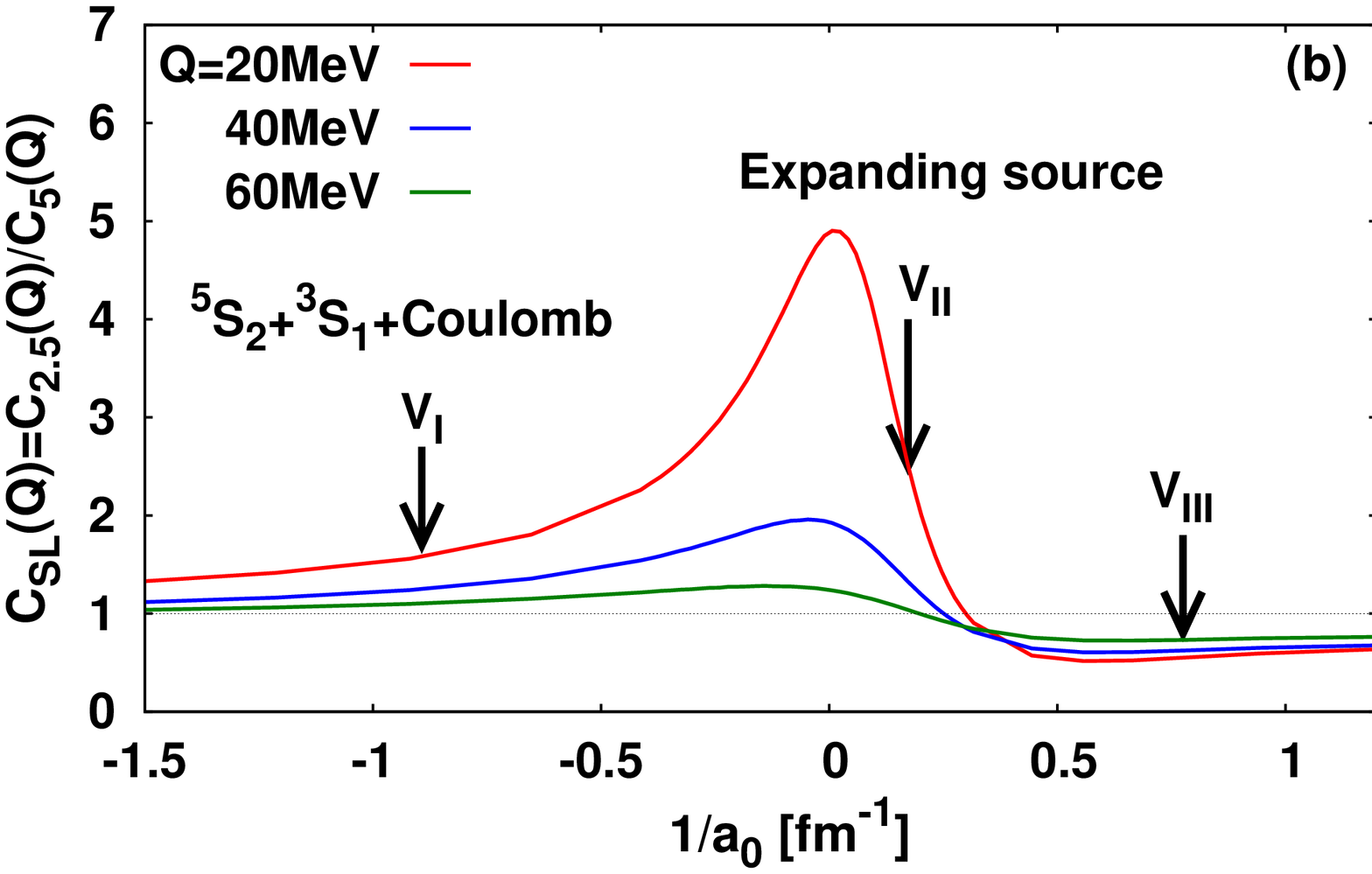} 
 \caption{(a) $C_{\rm SL}(Q)$ as a function of (a) $Q$ for
 three typical potentials  (b) $a_0^{-1}$.  In both figures,
 both the strong and Coulomb interactions are included. }
 \label{fig:mixedratio}
\end{figure}

Figure~\ref{fig:mixedratio}(a) demonstrates the effect of the dynamical
property  on $C_{\rm SL}(Q)$:  Its comparison to Fig.\ref{fig:csl} for
the static source indicates no significant difference
 as far as the ratio $C_{\rm SL}(Q)$ is concerned.
Figure~\ref{fig:mixedratio}(b) shows $C_{\rm SL}(Q)$ as a function of $a_0^{-1}$:
Its comparison to Fig.\ref{fig:c2}(b) on $C(Q)$ implies that
the strong $N\Omega$ interaction can be constrained by the measurements of this
ratio.
Moreover, taking the ratio of $C(Q)$ reduces the apparent reduction of its 
sensitivity to  the strong interaction due to the purity factor.
There are in principle two ways to extract $C_{\rm SL}(Q)$ experimentally
in ultrarelativistic heavy ion collisions at RHIC and LHC:  (i)
Comparison of the peripheral and central collisions for the same nuclear system,
 and (ii) comparison of the central collisions with
different system sizes, e.g. central Cu+Cu collisions and central Au+Au
collisions at RHIC.

\textit{Conclusion.\textemdash}
Motivated by the strong attraction at short distance between the proton
and the $\Omega$ baryon
in the spin-2 channel suggested by recent lattice QCD simulations, we studied the
 momentum correlation of  $p\Omega$ emission from relativistic heavy ion collisions.
Not only the elastic scattering in the spin-2 channel, but also the strong absorption 
in the spin-1 channel and the long-range Coulomb attraction are taken into account in 
our analysis.
Depending on the strength of the $p\Omega$ attraction, the correlation function at small
relative momentum changes substantially 
near the unitary limit.  We have proposed that the ratio of the correlation function
between the small and
large collision systems, $C_{\rm SL}(Q)$, is insensitive to the Coulomb interaction
and to the source model of the emission.
Thus  it provides a useful measure to extract the strong interaction part of the  $p\Omega$ attraction
from the experiments at RHIC and LHC.
Introduction of a realistic source model 
and  relativistic treatment of $C(Q)$
\cite{lednicky82:_influence} would be necessary for
more quantitative evaluation at RHIC and LHC energies, which are left for 
further studies in the near future.

\acknowledgements
We thank HAL QCD Coll. for
providing lattice data in Fig.1. 
This work was supported in part by the Grants-in-Aid for Scientific
Research on Innovative Areas from MEXT (Nos. 24105008 and 24105001) and 
Grants-in-Aid for Scientific Research from JSPS (Nos. 15K05079, 15H03663, 16K05349, 16K05350).  
 T.H. was supported in part by RIKEN iTHES Project.
K.M. was supported in part by the National Science Center, Poland under grant: Maestro DEC-2013/10/A/ST2/00106.

\end{document}